\documentclass[12pt,preprint]{aastex}

\newdimen\minuswidth    %define @ width of minus sign for tables
\setbox0=\hbox{$-$}
\minuswidth=\wd0
\catcode`@=\active
\def@{\kern\minuswidth}
\newdimen\digitwidth    %define ! a one digit width for tables
\setbox0=\hbox{\rm0}
\digitwidth=\wd0
\catcode`!=\active
\def!{\kern\digitwidth}

\def\reference#1\par{\parindent0pt\hangindent20pt\hangafter1 #1\par}

\received{}
\begin{document}
\shorttitle{Empirical Mass Loss Law}
\shortauthors{Origlia et al.}

\title{The First Empirical Mass Loss Law for Population II Giants$^1$}
\altaffiltext{1}{
This work is based on observations made with the Spitzer Space Telescope, 
which is operated by the Jet Propulsion Laboratory, 
California Institute of Technology under a contract with NASA.
Support for this work was provided by NASA through 
an award issued by JPL/Caltech.}

\author{Livia Origlia$^2$, Robert T. Rood$^3$, Sara Fabbri$^4$, 
Francesco R. Ferraro$^4$, 
Flavio Fusi Pecci$^2$, R. Michael Rich$^5$} 
\altaffiltext{2}{INAF--Osservatorio Astronomico di Bologna,
Via Ranzani 1, I--40127 Bologna, Italy, livia.origlia@oabo.inaf.it,
flavio.fusipecci@oabo.inaf.it}
\altaffiltext{3}{Astronomy Department, University of Virginia, Charlottesville, VA 22903,
rtr@virginia.edu}
\altaffiltext{4}
{Universit\`a degli Studi di Bologna, Dip. di Astronomia,
Via Ranzani 1, I--40127 Bologna, Italy,
sara.fabbri@studio.unibo.it,francesco.ferraro3@unibo.it}
\altaffiltext{5}{
Department of Physics and Astronomy, University of California
at Los Angeles, Los Angeles, CA 90095-1547,
rmr@astro.ucla.edu}
%\medskip

\begin{abstract}
Using the Spitzer IRAC camera we have 
obtained mid-IR photometry of the red giant branch stars 
in the Galactic globular cluster 47~Tuc. 
About  100
stars show an excess of mid-infrared light above that expected from
their photospheric emission. This is
plausibly due to dust formation in
mass flowing from these stars. 
This mass loss extends down to the level of the horizontal branch
and increases with luminosity. The mass loss is episodic, occurring in
only a fraction of stars at a given luminosity.  
Using a simple model and 
our observations we derive mass loss rates for these stars. 
Finally, we obtain
the first empirical mass loss formula calibrated with observations of
Population II stars. The dependence on luminosity of our mass loss rate is considerably shallower 
than the widely used Reimers Law. The results presented here are the
first from our Spitzer survey of a carefully chosen sample of 17
Galactic Globular Clusters, spanning the entire metallicity range from
about one hundredth up to almost solar.  
\end{abstract}


\section{Introduction}
\label{intro}

Though the current generation of theoretical models can reasonably
reproduce the general framework of stellar evolution, there are still
a number of physical phenomena which are poorly understood. Among
these the stellar mass loss (hereafter ML) is one of the most
vexing. This is particularly the case for cool stars. Indirect
evidence shows that ML strongly affects all of their late stages of
evolution, yet we have little theoretical or observational guidance of
how to incorporate ML into our models.  We have relied on empirical
laws like that of \citet{rei75a,rei75b} based on observations of
Population~I giants.  While subsequent work
\citep{mul78,gol79,jud91,cat00} led to slight refinements, a ``Reimers
Law'' or some variant has been the {\em only} basis for stellar
evolutionary models of cool stars at all ages and metallicities.
Indeed, a ML law directly calibrated on Population II low-mass giants
has never been determined. Galactic globular clusters (GGCs) are particularly attractive
observational targets because there is so much indirect, {\em but
quantitative,} evidence for ML. This includes the observed morphology
of the horizontal branch (HB) in the cluster color-magnitude diagrams
(CMDs), the pulsational properties of the RR Lyrae stars, and the
absence of asymptotic giant branch (AGB) stars brighter than the red
giant branch (RGB) tip, and the masses inferred for white dwarfs in
GCCs \citep{roo73,ffp75,ffp76,ren77,ffp93,cru96,han05,kal07}. 

\begin{figure*}
%\epsscale{1.8}
\plotone{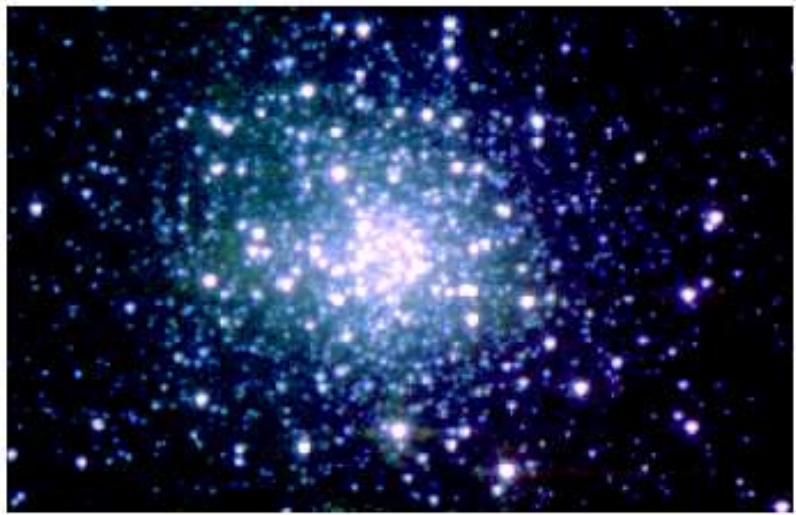}
\caption{Three color (3.6$\mu$m (blue), 6$\mu$m (green), 8$\mu$m (red) mosaiced image of 47~Tuc 
from IRAC. 
\label{ima}}
\end{figure*}

\begin{figure*}
%\epsscale{2.3}
\plottwo{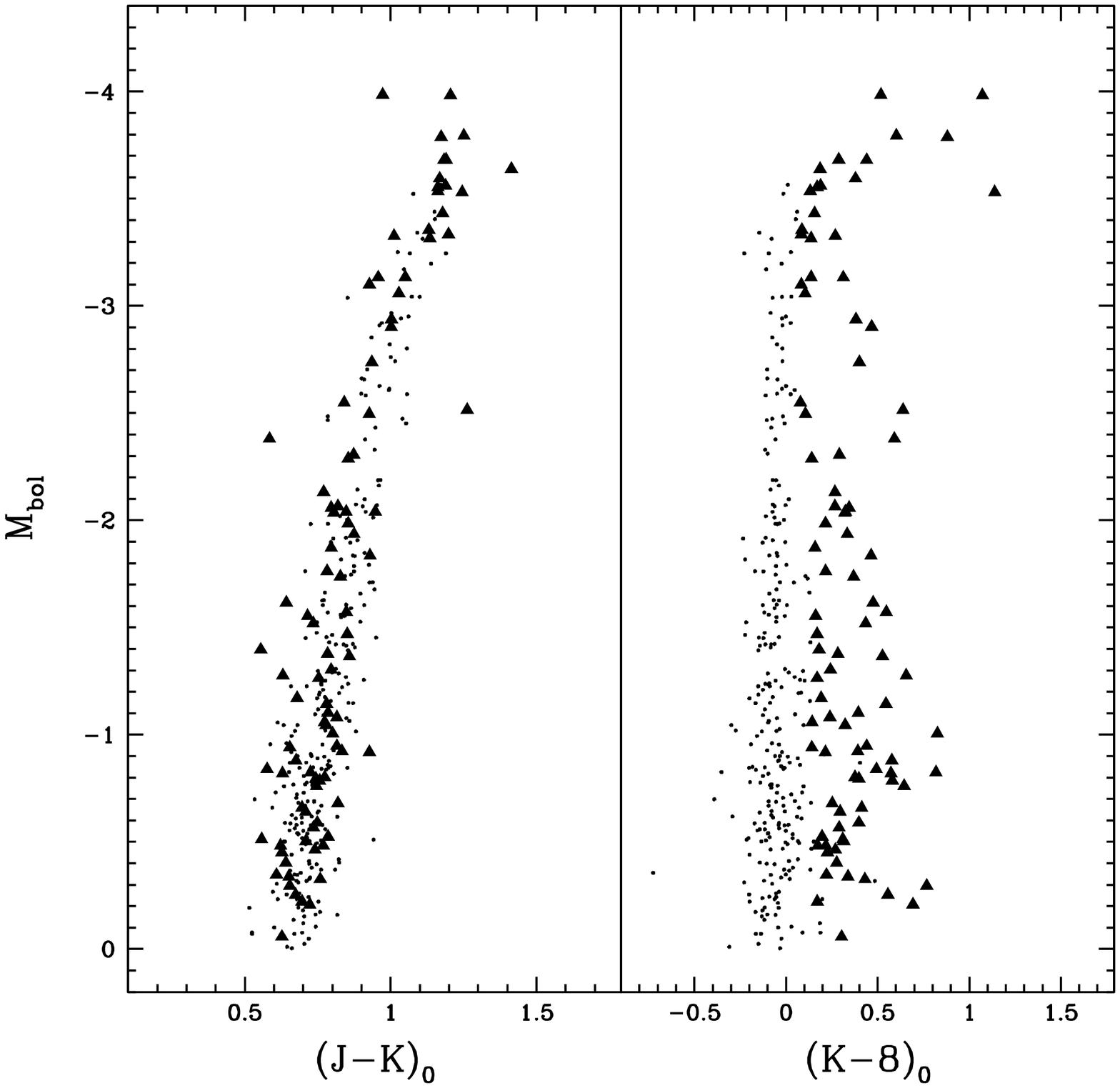}{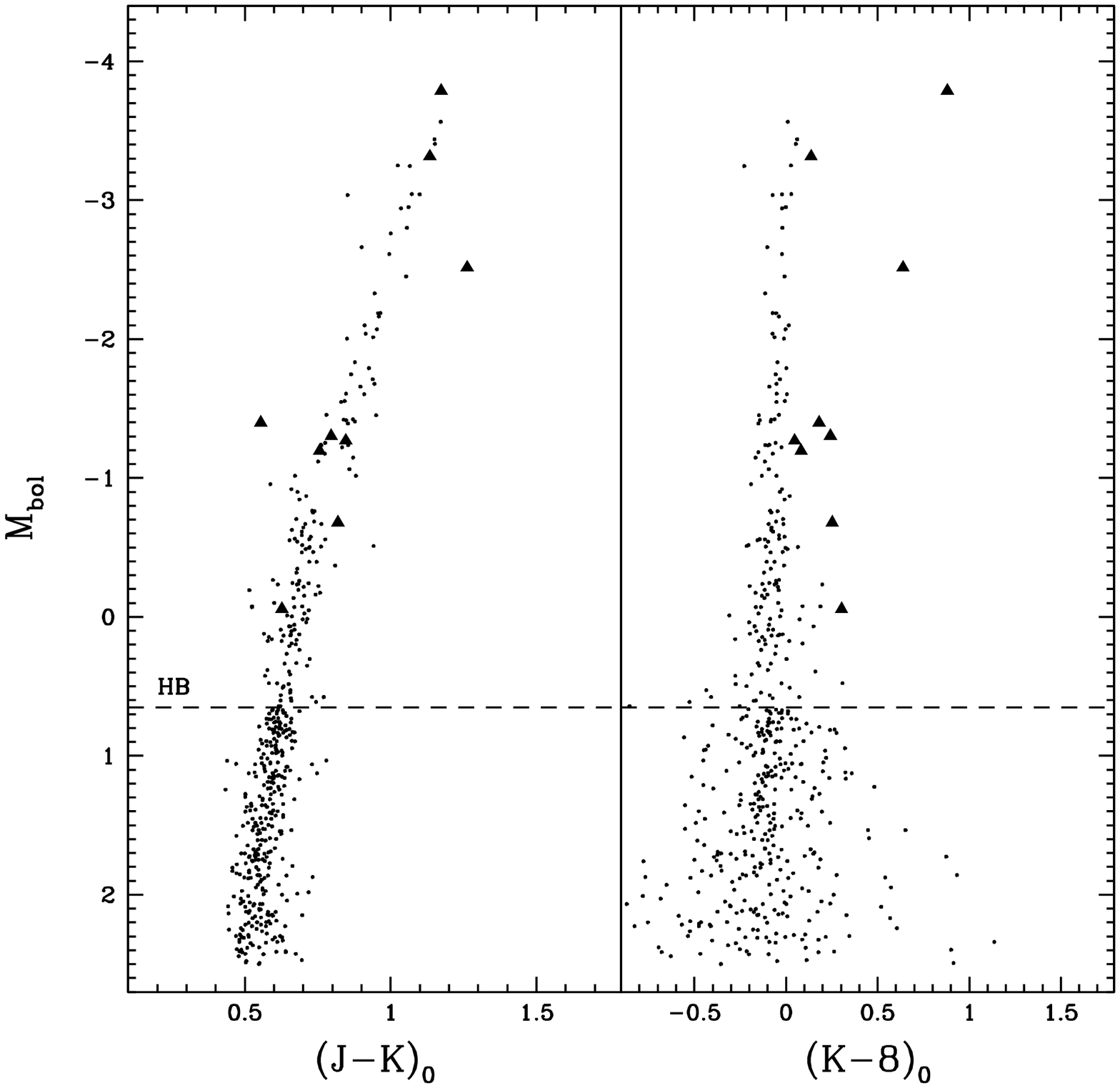}
\caption{Absolute CMDs in the near IR and Spitzer planes of 47~Tuc. 
Left panel: stars detected in the shallow survey down  
to $K\approx11$ or $M_{\rm bol}\approx 0$ over a useful
$8'\times5'$ field of view.
Right panel: stars detected in the deeper survey down to $K\approx14$ which
excludes the inner $\approx2'$ (by radius) region.
The dashed horizontal line marks the position of the HB level.  
Stars with ($K-8$)$_0$ color excess at a 3$\sigma $ level are marked as filled triangles.
\label{cmd}}
\end{figure*}

Direct evidence for ML in GGC RGB stars has been inferred from blue
shifted features of photospheric lines like H$\alpha$
\citep{dup86,mau06}.  These are difficult observations and hard to
convert to ML rates.  One can also observe the outflowing gas much
further from the star after it has formed dust.  We detected this
circumstellar (CS) matter using ISOCAM and were able to derive ML
rates for a modest sample of RGB stars \citep{ori02}.  Mid-IR
observations have the advantage of sampling an outflowing gas fairly
far from the star (typically, tens/hundreds stellar radii).  Such gas
typically left the star a few decades previously, hence the inferred
ML rate is also smoothed over such a timescale.  From a combined
physical and statistical analysis, our ISOCAM study provided ML rates and frequency.
We found that {\it i)}~the largest ML occurs near the very end of
the RGB evolutionary stage and is episodic, {\it ii)} typical rates
are in the range $10^{-7} < dM/dt < 10^{-6} M_{\odot}\,{\rm yr^{-1}}$,
{\it iii)} ~the modulation timescales must be greater than a few
decades and less than a million years, and {\it iv)}~there is evidence
for dusty shells at even the lowest metallicities. However, our ISOCAM
survey suffered from two significant limitations. The small
sample of observed clusters and the consequent low number of detected 
giants allowed us to reach only weak conclusions on the ML dependence
on metallicity and HB morphology. Further, the modest spatial
resolution and sensitivity compromised our ability to measure lower ML
rates near the RGB tip and made it impossible to explore ML much below the
RGB tip.

\section{The Spitzer IRAC survey}

A sample of 17 massive GGCs, 4--5 per each 0.5 dex bin in
metallicity between ${\rm [Fe/H]} =-2.3$ and $-0.5$ and different HB
morphologies within each bin has been observed with IRAC onboard
Spitzer with 26 hr of observing time allocated to our program (ID
\#20298) in Cycle~2.  For all these clusters complementary near-IR and
UV photometry are available to properly characterize both the red and
the blue sequences.

In this Letter we discuss the first cluster analyzed, 47~Tuc.  This
cluster was observed on September 21, 2005.  We used a $1\times3$ grid
with small cycling dither pattern to cover the central $9' \times 5'$
of the cluster in all the four IRAC filters and the {\it High Dynamic
Range} Readout Mode, to avoid saturation of the brightest giants.  A
total observing time of 1.2\,hr allowed us to reach $K\le14$ with
$S/N\approx20$.  
Fig.~\ref{ima} shows the mosaiced three-color image of 47~Tuc.
The Post BCD mosaic frames from the Spitzer Pipeline
(Software Version: S13.2.0) have been photometrically reduced with
ROMAFOT \citep{buo83}, a software package optimized for Point Spread
Function fitting in crowded and undersampled stellar fields.  Because
of the intrinsic luminosity of 47~Tuc, the crowding in its central
region is critical even at 8~$\mu$m.  We thus obtained two photometric
catalogs: a shallow one down to $K\approx11$ over a useful
$8'\times5'$ field of view and a deeper one down to $K\approx14$ which
excludes the inner $\approx2'$ (by radius) region.  The overall
photometric uncertainty in all the four IRAC filters is $\le0.1$ mag.
Complementary ground-based near-IR photometry of the central region at
high spatial resolution has been obtained using IRAC2\@ ESO2.2m, and
SOFI@ESO-NTT \citep{fer00,val04} and supplemented with 2MASS data in
the external region.  
The degree of completeness of the near-IR survey is
100\% over the full magnitude range covered by the Spitzer survey.
The shallow Spitzer catalog counts almost 400 stars and it is $\ge 86\%$ complete in the upper two RGB
magnitudes, and $\approx66\%$ complete at $0\ge M_{\rm bol}>-2$, while the deeper
Spitzer catalog in the outer region counts almost 500 stars and it is $\ge$83\% complete down to $K\le14$.

The two Spitzer catalogs have been combined
and placed onto the 2MASS astrometric system by cross-correlating the
Spitzer and the ground-based near-IR surveys.  The final catalog
contains almost 800 stars with  $J,~K$, and Spitzer photometry in each of the
four filters.  

The dereddened $K_0$ magnitudes and $(J-K)_0$ colors have been also
used to compute the stellar bolometric magnitudes and effective
temperatures, by using the transformation of \citet{mon98},
reddening ($E(B-V)=0.04$) and distance modulus ($(m-M)_0=13.32$) from
\citet{fer99}.  Fig.~\ref{cmd} shows the absolute color-magnitude
diagrams (CMDs) of 47~Tuc in the near and mid IR planes. 
By knowing the stellar temperature and bolometric magnitude and using a
typical RGB stellar mass of 0.8--0.9 M$_{\odot}$,
we finally estimate the stellar radius and gravity 
from the following equations:
\begin{displaymath}
L_{\star} = 4\pi R_{\star}^2 \times \sigma T_{\rm eff}^4 ~~~~~g = G M_{\star}/R_{\star}^2
\end{displaymath}

\begin{figure*}
%\epsscale{2.0}
\plotone{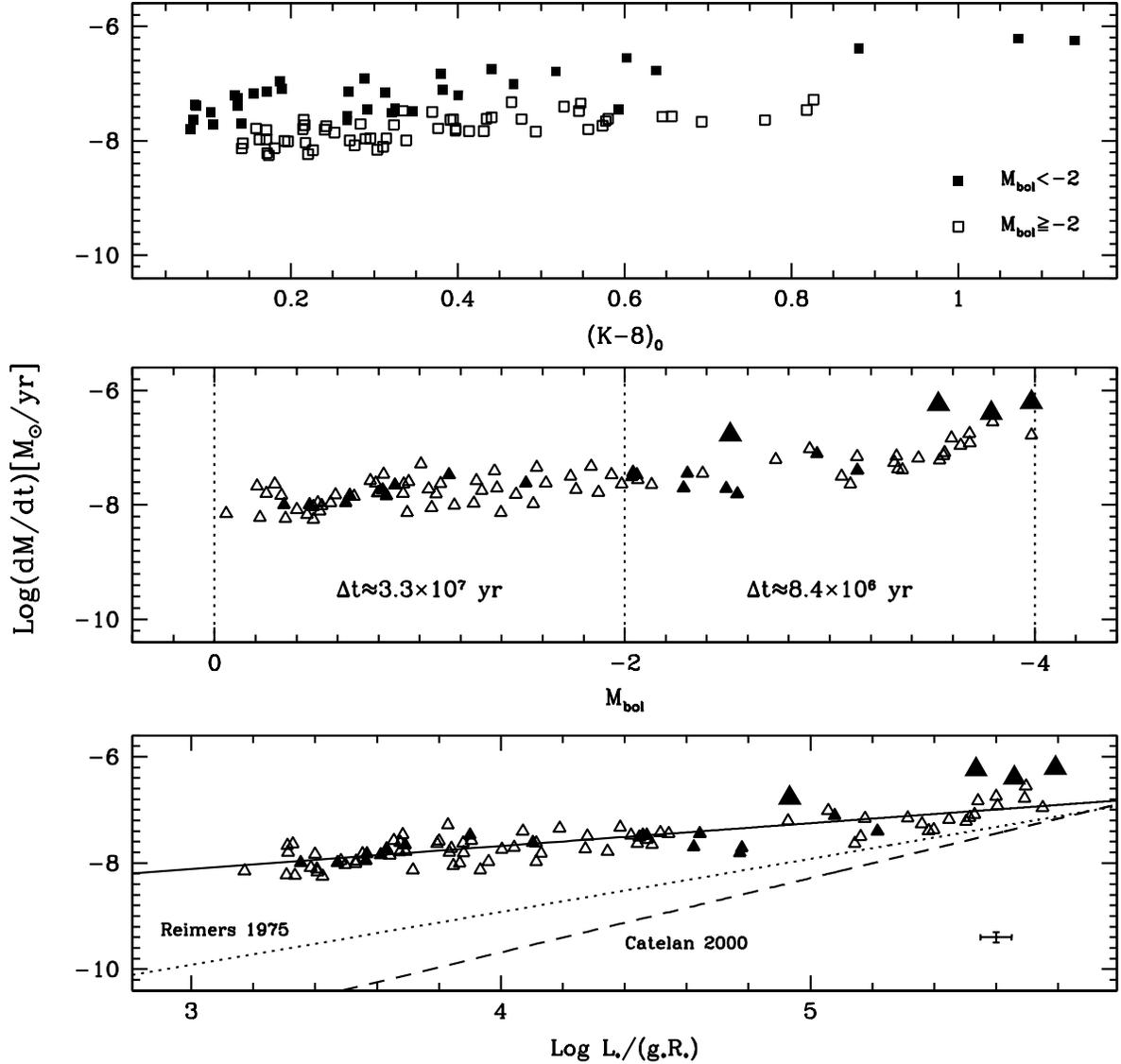}
\caption{ML rates for the Spitzer sources with dust excess, as a
function of the observed ($K-8$)$_0$ color (top panel), bolometric
magnitude (middle panel) and normalized stellar luminosity (bottom
panel).  In the top panel stars with $M_{\rm bol}<-2$ and $M_{\rm bol}\ge-2$ are
plotted as filled and open squares, respectively.  In the middle and
lower panels
filled big triangles mark those giants which are known long period
variables, filled small triangles are other AGB stars.  In the middle panel
the evolutionary timescale $\Delta t$ in 2 luminosity
intervals are also reported.  In the bottom panel
all of the AGB stars are excluded
from the fit (solid lines). 
Typical random error bars are shown in the bottom right corner.
The empirical laws by
\citet{rei75a,rei75b} with $\eta =0.3$ (short-dashed line) and \citet{cat00}
(long-dashed line) calibrated on Population~I giants are also shown
for comparison.
\label{ml}}
\end{figure*}

\section{Dust excess and mass loss rates}

As CS dust condenses in an outflowing wind, it can be detected as a
mid-IR excess.  In our pilot project using ISOCAM photometry in the 10 $\mu $m
window and  assuming a $\nu B_{\nu}$ emissivity, we showed that the
bulk of CS dust around the RGB tip giants typically has temperatures
in the range 300--500\,K.  The IRAC bands between 3.6 and 8 $\mu $m
can also be used to detect this warm dust when coupled with near IR
photometry used to properly characterize the stellar counterpart.

In order to select candidate stars with dust excess, we define first
the mean ridge lines in each of the $K_0,~(K-IRAC)_0$ CMDs to define
the average color of the stars with purely photospheric emission and
to determine the overall photometric errors ($\sigma$) at different magnitude
bins.  These $(K-IRAC)_0$ colors are $\approx 0.0 \pm 0.1$
along the entire RGB range sampled by our survey. This is in good
agreement with the prediction of theoretical model atmospheres with
$T_{\rm eff}=3500$--5000 K from the Kurucz database.  Since the 8 $\mu
$m IRAC band is the most sensitive to warm dust emission, stars are
flagged as dusty if they show a $(K-8)_0$ color excess $\ge
n\sigma$. These stars are also the reddest in the other IRAC bands.
For these stars, we 
also compute the pure dust emission at 8 $\mu m$, by subtracting 
from the the total observed flux  
the photospheric contribution, given by $F_8^{\rm phot} = 
F_8^{\rm Vega}\times 10.^{[-0.4\times (K-(K-8)_{\rm phot})]}$, where $(K-8)_{\rm phot}$ 
is the mean ridge line color without dust excess.

The number of giants with dust excess in the
shallow survey of 47~Tuc is 93 at a 3$\sigma$ level.  In the deeper
survey of the cluster outer region, no stars with dust excess have
been detected below the HB level.  The seven 47~Tuc stars which showed
dust excess in our ISOCAM survey have been also detected by Spitzer
and confirmed as dusty stars.  Among the Spitzer dusty giants, four
are known long period variables (V1, V4, V6, \& V8 in \citet{cle01})
and an other 17 stars are classified as AGB stars by
\citet{bec06}.  Hence, in the following we classify the remaining 74
giants as true RGB dusty stars.

In order to obtain the ML rates we use our modified version of the
DUSTY code \citep{ive99,eli01}, to compute the emerging
spectrum and dust emission at the IRAC wavelengths.  We adopt Kurucz
model atmospheres for the heating source and for the dust a mixture of
warm silicates with an average grain radius $a=0.1\,\mu $m.  Slightly
different choices for the latter two parameters have negligible impact
in the resulting IRAC colors and mass loss rates.  While radiation
pressure acting on dust might plausibly drive winds in luminous, metal
rich red giants \citep{wil00}, the GGC stars are generally neither
luminous nor metal rich enough for this mechanism to be efficient.
Hence we run the DUSTY code under the general assumption of an
expanding envelope at constant velocity $v_{\rm exp}$ with a density
profile $\eta \propto r^{-2}$, a dust temperature for the inner
boundary $r_{\rm in}$ of 1000~K and a shell outer boundary $
r_{\rm out}=1000 \times r_{\rm in}$.  We then computed a large grid of DUSTY
models with stellar temperatures in the 3500--5000 K range and optical
depths at 8 $\mu $m ($\tau_8$) between 10$^{-5}$ and 10$^{-1}$.
For each star with dust excess, we enter the grid with its empirical
stellar temperature and $(K-IRAC)_0$ colors, and we exit with
prediction for the optical depth, emerging flux, dust fractional contribution 
and envelope radius.
The mass loss rates are computed by using the formula
\begin{displaymath}
dM/dt = 4 \pi  r_{\rm out}^2 \times \rho_{\rm dust}
\times v_{\rm exp} \times \delta
\end{displaymath}
where 
\begin{displaymath}
\rho_{\rm dust}\propto
\rho_{\rm g} \tau_8 F_8({\rm obs})/F_8({\rm mod}) D^2 
\end{displaymath}

\noindent
is the dust density, $\rho_{\rm g}=3\,
{\rm g\,cm^{-3}}$ is the grain density, 
$F_8(obs)$ and $F_8(mod)$ are the observed and model dust emission at 8 micron,
respectively,
$D$ the distance and $\delta$ the gas to dust ratio.
A lower limit to $\delta$ is given by $1/Z$ where $Z$ is the global
metallicity.  $v_{\rm exp}$ is a free parameter, which should scale
like $\delta ^{-0.5}$ if dust and gas are coupled.  If 
the number of grains is increased (by decreasing $\delta$), the momentum per
grain is shared with fewer gas molecules, thus implying an enhancement
of $v_{\rm exp}$ \citep{hab94,van00}.  
For 47~Tuc we adopt
$\delta\approx 1/Z\approx 200$ and $v_{\rm exp}=10~{\rm km\,s^{-1}}$. 
The latter is the average expansion velocity measured in luminous, low mass giants
\citep[see e.g.][]{net93} which ranges between a few and $\approx 20~{\rm km s^{-1}}$. 

\section{Results and Discussion}         

For the dusty stars in 47~Tuc sampled by our Spitzer survey,
Fig.~\ref{ml} shows the inferred mass loss rates as a function of {\it
i)}~ the observed $(K-8)_0$ color, {\it ii)}~the bolometric magnitude
and {\it iii)}~the normalized luminosity.  The mass loss rate
increases with increasing color excess and stellar luminosity.  
Also,
the bulk of the mass loss along the RGB should occur above the HB
level.  
The provisional empirical law based on such a first set of
observations gives:
\begin{eqnarray*}
dM/dt = 
& C \times 4 \times 10^{-10} \times (L/gR)_{\odot}^{0.4}  
\end{eqnarray*}
where
\begin{eqnarray*}
C = & (\delta/200)^{0.5}\times (v_{\rm
exp}/10)\times(\rho_{\rm g}/3) 
\end{eqnarray*}
and 
$L_{\odot}$, $g_{\odot}$, and $R_{\odot}$ are the stellar luminosity, 
gravity and radius in solar units. In this study of 47~Tuc $C=1$. 
Only true RGB stars are used to derive the fitting formula.
Errors are as follows: 
$\approx 10$\% for 
$L/gR$ and the fit exponent  
and $\approx25$\% for the fit zero point. The latter 
is also a good estimate of the average random error on the final 
mass loss rates.   
As shown in Fig.~\ref{ml}, this new law calibrated on Population II
RGB stars is significantly flatter than the original Reimers
formulation and the one revised by \citet{cat00}, which have slopes of
1 and 1.4, respectively.  

As we found in our ISOCAM survey, only a
fraction of stars along the RGB are currently losing mass: $\approx
32$\% in the upper 2 mag, $\approx16$\% down to the HB.  For this
reason we conclude that ML is episodic. Basically mass loss is ``on''
for only some fraction of the time, $f_{\rm on}$. By using a suitable
evolutionary track for a RGB star of $M=0.9\, M_{\odot}$ and $Z=0.004$
\citep{pie06}, we can derive the evolutionary timescale in each
luminosity interval (see Fig.~\ref{ml}).  Multiplying this by $f_{\rm
on}$, we find that the timescale ML is ``on'' is less than a few Myr
in each interval.  
By using the simple equation 
\begin{eqnarray*}
\Delta M_{\rm RGB}  = \Sigma_i~ ( dM/dt_i \times \Delta t_i \times f_{{\rm on}_i} )
\end{eqnarray*}
to integrate the mass loss formula multiplied by
$f_{\rm on}$ over the RGB evolution time $\Delta t$, 
we find that the total mass
lost on the RGB is $\Delta M_{\rm RGB}\approx 0.23\pm 0.07 M_{\odot}$.
 
In forthcoming papers we will present the results for the other
clusters, with the ultimate goal of providing {the first empirical ML
law for Population~II stars} calibrated over a large range of
metallicity, and investigating whether observed ML in individual stars
within a cluster correlates with that cluster's HB morphology and 
if ML itself is involved in the second-parameter problem.

\acknowledgments
This research was supported by 
the Ministero dell'Istru\-zio\-ne, Universit\`a e Ricerca (MIUR), the 
PRIN-INAF 2006 and the Italian Space Agency (ASI).
RTR and RMR acknowledge support from Spitzer Science Center Grant GO-20298.
We warmly thanks Elena Valenti for having provided the SOFI near IR photometric catalog 
and the anonymous Referee for his/her useful comments.

\end{document}